\documentclass{article}

\usepackage{arxiv}

\usepackage[utf8]{inputenc} % allow utf-8 input
\usepackage[T1]{fontenc}    % use 8-bit T1 fonts
\usepackage{hyperref}       % hyperlinks
\usepackage{url}            % simple URL typesetting
\usepackage{booktabs}       % professional-quality tables
\usepackage{amsfonts}       % blackboard math symbols
\usepackage{nicefrac}       % compact symbols for 1/2, etc.
\usepackage{microtype}      % microtypography
\usepackage{lipsum}
\usepackage{amsmath, amsfonts, amsthm, amssymb, mathtools} % math symbols
\usepackage{physics} % neatly formatted vectors and derivatives
\usepackage{cite} % neatly formatted citations

\usepackage{pgfplots}
\usetikzlibrary{arrows}

\newcommand{\Sof}[1]{S\left[\ #1\ \right]}
\newcommand{\Tof}[1]{T\left[\ #1\ \right]}

\theoremstyle{definition}
\newtheorem{definition}{Definition}

\title{A Commentary on the Linearity and Time-Invariance of ODE-Based Systems}

\author{
  Parker S.~Ruth\thanks{\href{https://parkersruth.github.io/}{\texttt{parkersruth.github.io}}}\\
  Department of Bioengineering and\\ Paul G.~Allen School of Computer Science and Engineering \\
  University of Washington\\
  185 Stevens Way, Seattle, WA, 98195 \\
  \texttt{psr23@uw.edu} \\
   \And
 Herbert M.~Sauro\thanks{\href{http://sys-bio.org/}{\texttt{sys-bio.org}}}\\
  Department of Bioengineering\\
  University of Washington\\
  185 Stevens Way, Seattle, WA, 98195 \\
  \texttt{hsauro@uw.edu} \\
}

\begin{document}
\maketitle

% ABSTRACT
\begin{abstract}
Linear time-invariant (LTI) systems appear frequently in natural sciences and engineering contexts. Many LTI systems are described by ordinary differential equations (ODEs). For example, biological gene regulation, analog filter circuits, and simple mechanical, electrical, and hydraulic systems can all be described with varying approximations as LTI systems using ODEs. While linearity and time-invariance are straightforward to demonstrate for closed-form system definitions, determining whether an ODE describes a system with LTI properties is less obvious and rarely discussed in depth in the literature. Complications arise due to slightly different definitions of linearity in different contexts. This commentary is intended to provide clarity on this subtle point, and act as an instructional aid or educational supplement.
\end{abstract}

% keywords can be removed

\section{Systems Without Feedback}

Let $S$ be an arbitrary system that maps a real-valued time-varying input $x(t)$ to a real-valued time-varying output $y(t)$. In the simplest case, the output is a deterministic function of the input so that $y(t) = \Sof{x(t)}$.
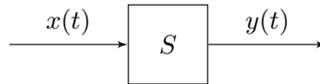
\begin{center}
\begin{tikzpicture}[node distance=2.5cm,auto,>=latex']
    \node [minimum size=3em, draw] (a) {$S$};
    \node (b) [left of=a,node distance=6em, coordinate] {$x(t)$};
    \node (c) [right of=a,node distance=6em, coordinate] {$y(t)$};
    \path[->] (b) edge node {$x(t)$} (a);
    \path[->] (a) edge node {$y(t)$} (c);
\end{tikzpicture}
\end{center}
The input $x(t)$ is assumed to be known and controlled (perhaps by an experimenter or control system).
When convenient, we may omit the time parameter without changing the meaning, as in $ \Sof{x} = y $.

\begin{definition}
The system without feedback $\Sof{x(t)}$ is linear iff
\begin{gather}
    \Sof{x_1(t)} = y_1(t) \qquad\text{and}\qquad \Sof{x_2(t)} = y_2(t) \nonumber \\[0.5em]
    \text{implies that} \nonumber \\[0.5em]
    \Sof{\alpha x_1(t) + \beta x_2(t)} = \alpha y_1(t) + \beta y_2(t) \label{linear}
\end{gather}
The system without feedback $\Sof{x(t)}$ is time-invariant iff
\begin{gather}
    \Sof{x(t)} = y(t) \nonumber \\[0.5em]
    \text{implies that} \nonumber \\[0.5em]
    \Sof{ x(t + \delta) } = y(t + \delta) \label{time-invariant}
\end{gather}
A system is linear time-invariant (LTI) iff the system is both linear and time-invariant \cite{ogata_modern_1970, lathi_signal_1998, sauro_control_2019}. A simple but important corollary of (\ref{linear}) is that $\Sof{0} = 0$ for any LTI system $S$.
\end{definition}

Determining linearity is straightforward when there is no feedback. For example, consider the system
$$y = \Sof{x} = ax $$\\[-0.5em]
We can confirm mechanically that $S$ is LTI by checking that (\ref{linear}) and (\ref{time-invariant}) are satisfied. Let $\alpha$ and $\beta$ be arbitrary constants, and let $x_1(t)$ and $x_2(t)$ be arbitrary inputs.
$$\Sof{\alpha x_1 + \beta x_2} = a(\alpha x_1 + \beta x_2) = a \alpha x_1 + a \beta x_2 = \alpha y_1 + \beta y_2$$
$$ \Sof{x(t + \delta)} = ax(t + \delta) = y(t + \delta) $$

Furthermore, a system can be shown to be non-LTI by finding a single counterexample to (\ref{linear}) or (\ref{time-invariant}). For example, consider the system
\begin{equation} y = \Sof{x} = ax + b \label{notlti} \end{equation}
This system is not LTI because for $x_1(t) = 3$ and $x_2(t) = 4$, we find that
$$ \Sof{x_1 + x_2} = a(x_1 + x_2) + b = a(3 + 4) + b = 7a + b $$
which, in violation of (\ref{linear}) does not equal
$$ \Sof{x_1} + \Sof{x_2} = a(x_1) + b + a(x_2) + b = 3a + b + 4a + b = 7a + 2b $$
Note that although $y = ax + b$ is frequently referred to as ``linear'' in other contexts because it describes a line with intercept $b$, strictly speaking this is an affine transformation; it is \emph{not} a linear system by our definition.

\section{Systems With Feedback}

It is also possible for the system to exhibit some feedback, so that the output depends on both the input and the output itself $y(t) = \Sof{x(t), y(t)}$.
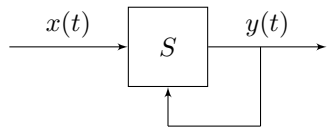
\begin{center}
\begin{tikzpicture}[node distance=2.5cm, auto, >=latex']
    \node [minimum size=3em, draw] (a) {$S$};
    \node (b) [left of=a,node distance=6em, coordinate] {$x(t)$};
    \node (c) [right of=a,node distance=6em, coordinate] {$y(t)$};
    \path[->] (b) edge node {$x(t)$} (a);
    \path[->] (a) edge node {$y(t)$} (c);
    \node [right of=a, node distance=3.5em, coordinate] (ac) {yo};
    \node (fb) [below of=ac, node distance=3em, coordinate] {};
    \path[-] (ac) edge node {} (fb);
    \node (fbb) [left of=fb, node distance=3.5em, coordinate] {};
    \path[-] (fb) edge node {} (fbb);
    \path[->] (fbb) edge node {} (a);
\end{tikzpicture}
\end{center}
It is no longer immediately clear how to apply \eqref{linear} and \eqref{time-invariant} to test if $S$ is LTI. One solution is to ``unroll'' the feedback to obtain a closed-form expression for $y$ in terms of only x. For example, consider the following system with feedback:
$$y(t) = \Sof{x(t), y(t))} = a y(t) + b x(t)$$
We can test if the system is LTI by computing an equivalent system $T$ so that $y(t) = \Tof{x(t)}$. Solving the equation above for $y(t)$ we now obtain:
$$y(t) = \Tof{x(t)} = \frac{b}{1-a} x(t)$$
Clearly $S$ and $T$ have the same behavior since the relationship between $y$ and $x$ is identical for all times. Since we can easily confirm that $T$ is LTI by applying \eqref{linear} and \eqref{time-invariant}, we can also say that $S$ is LTI.

However, this strategy of ``unrolling'' a system with feedback is not always a simple algebraic manipulation. Consider, for example, the following system with feedback:
\begin{equation} y(t) = \Sof{x(t), y(t))} = a \dv{y}{t} (t) + b x(t) \label{simple-ode} \end{equation}
Applying the ``unrolling'' trick now requires solving this ordinary differential equation (ODE). The general solution is
\begin{equation} y(t) = \Tof{x(t)} = e^{\frac{t}{a}} \left( y_0 - \tfrac{b}{a} \int_0^t e^{-\frac{\tau}{a}} x(\tau) \dd\tau \right)
\label{simple-ode-sol} \end{equation}
where $y_0 = y(0)$ is the initial condition of the system. However, the explicit reliance on the initial condition becomes problematic when we try to apply Definition 1. The parameter $y_0$ produces an entire family of different ``unrolled'' systems that satisfy the original equation \eqref{simple-ode}. We now wish to determine if any members of this family are LTI.

To satisfy linearity, recall that every LTI system must have $S[0] = 0$. Applying this constraint to $T$, we find that
$$\Tof{0} = y_0 e^{at} = 0$$
which is only true when $y_0 = 0$. This leaves us with the following specific solution to \eqref{simple-ode} which now has no parameters.
$$\Tof{x(t)} = - \tfrac{b}{a} e^{at} \int_0^t e^{-a\tau} x(\tau) \dd\tau$$
It is easy to show that this solution is linear; however, we find that it fails the time-invariance test, since
$$ \Tof{x(t+\delta)} = - \tfrac{b}{a} e^{at} \int_0^t e^{-a\tau} x(\tau+\delta) \dd\tau \neq -\tfrac{b}{a} e^{a(t+\delta)} \int_0^{t+\delta} e^{-a\tau} x(\tau) \dd\tau =\Tof{x(t)}(t+\delta)$$
Therefore, there does not exist a particular closed-form solution to \eqref{simple-ode} that satisfies both \eqref{linear} and \eqref{time-invariant}. In particular, the linearity constraint requires us to set the initial conditions to zero; however, fixing a particular initial condition violates the requirement of time-invariance. To preserve time-invariance, we would need to both shift $x(t)$ and $y(t)$ in time, and also appropriately update our choice of $y_0$; however, doing so results in a different formula for $T$. Does this mean that system \eqref{simple-ode} is not an LTI system? To the contrary, we will find that system \eqref{simple-ode} in fact \emph{is} an LTI system; we have simply failed to find an LTI transformation $T$ \emph{without feedback} that is equivalent to the system $S$ \emph{with feedback}. Although the ``unrolling'' strategy has not succeeded, we would still like to determine whether $S$ is LTI.  Thus, we can extend the definitions of linearity and time-invariance to better accommodate systems with feedback.

\begin{definition}
The system with feedback $\Sof{x, y}$ is linear iff
\begin{gather}
    \Sof{x_1, y_1} = y_1 \qquad\text{and}\qquad \Sof{x_2, y_2} = y_2 \nonumber \\[0.5em]
    \text{implies that} \nonumber \\[0.5em]
    \Sof{\alpha x_1 + \beta x_2,\ \alpha y_1 + \beta y_2} = \alpha y_1 + \beta y_2 \label{linear-fb}
\end{gather}
The system with feedback $\Sof{x, y}$ is time-invariant iff
\begin{gather}
    \Sof{x(t), y(t)} = y(t) \nonumber \\[0.5em]
    \text{implies that} \nonumber \\[0.5em]
    \Sof{ x(t + \delta), y(t + \delta) } = y(t + \delta) \label{time-invariant-fb}
\end{gather}
\end{definition}

Using this new definition, it is straightforward to show that system \eqref{simple-ode} is LTI. Suppose that there exist two arbitrary solutions to the ODE:
$$ y_1 = a \dv{y_1}{t} + b x_1 \qquad\text{and}\qquad y_2 = a \dv{y_2}{t} + b x_2$$
Applying \eqref{linear-fb} directly, we obtain
\begin{align*}
    \Sof{\alpha x_1 + \beta x_2,\ \alpha y_1 + \beta y_2}
        &= a \dv{t} \left(\alpha y_1 + \beta y_2\right) + b \left(\alpha x_1 + \beta x_2\right) \\
        &= \alpha \left(a\dv{y_1}{t} + bx_1\right) + \beta \left(a\dv{y_1}{t} + bx_2\right) \\
        &= \alpha y_1 + \beta y_2 \label{linear-fb}
\end{align*}
which confirms that \eqref{simple-ode} is linear. Now applying \eqref{time-invariant-fb}, we find
$$ \Sof{x(t+\delta), y(t+\delta)} = a \dv{y}{t} (t+\delta) + b x(t + \delta) = y(t+\delta) $$
which confirms that \eqref{simple-ode} is time-invariant. In this way, we have shown that the ODE is LTI without needing to solve it.

We can likewise demonstrate that a system with feedback is not LTI by finding a counterexample to either \eqref{linear-fb} or \eqref{time-invariant-fb}. Consider the following system with feedback:
$$ y = \Sof{x, y} = \dv{y}{t} + x + a $$
where $a$ is an arbitrary constant. This system is not linear, since
$$ \Sof{\alpha x, \alpha y} = \alpha \dv{y}{t} + \alpha x + a \neq \alpha \left( \dv{y}{t} + x + a \right) = \alpha y $$

\section{Linearity and Time-Invariance of ODE-Based Systems}

By the new definition, we can now state more generally that any system written as a linear combination of derivatives of the input and output is LTI. Consider an arbitrary system with the following form.
$$ \Sof{x, y} = a_1 \dv{y}{t} + a_2 \dv[2]{y}{t} + \ldots + a_n \dv[n]{y}{t} + b_0 x + b_1 \dv{x}{t} + \ldots b_m \dv[m]{x}{t} = \sum_{i=1}^n a_i \dv[i]{y}{t} + \sum_{j=0}^m b_j \dv[j]{x}{t} $$
In an analysis similar to that above, it follows that \eqref{linear-fb} and \eqref{time-invariant-fb} both hold for this system; thus, it is LTI.

There may be multiple equivalent representations for a single system with feedback. For example, the following two systems are equivalent:
\begin{align*}
y &= \Sof{x, y} = \dv{y}{t} + \int y \dd{t} + x\\
y &= \Sof{x, y} = -\dv[2]{y}{t} + \dv{y}{t} - \dv{x}{t}
\end{align*}
The equivalence follows by differentiating the top system and solving for $y(t)$. In this way, any system described by a linear combination of derivatives and integrals of $y$ can be re-written to only contain derivatives terms. However, this does not work in general for systems with integrals of $x$. For example, consider the following.
\begin{equation} y = \Sof{x, y} = \dv{y}{t} + \int x \dd{t} \label{untangleable} \end{equation}
Eliminating the integral from this equation eliminates the zeroth order term in $y$, preventing us from being able to solve the equation for $y$. Instead, the best we can do is this:
\begin{equation} \dv{y}{t} = \dv[2]{y}{t} + x \label{untangled} \end{equation}
By the linearity of differentiation and integration, we can show that \eqref{untangleable} is LTI. However, when we re-write it in the form of an ODE, we cannot solve for $y$ to write out $\Sof{x, y}$ as we have done before. Nonetheless, we can still understand \eqref{untangled} to describe an LTI system even though it is not written with $y$ separated on its own. By allowing ourselves this flexibility in notation, we can now claim generally that {\bf every linear ODE describes an LTI system}. Indeed, a wide array of LTI systems with scientific and engineering applications are described in the form of ODEs.

\section{Conclusion}

We have presented an extension to the definition of linear time-invariance that accommodates systems with feedback. We presented methods for proving and disproving the linear time-invariance of systems with and without feedback. Finally, we showed that every linear ODE describes an LTI system.

\bibliographystyle{unsrt}  
\bibliography{references}

\end{document}